\documentclass{pasj00}
\draft

\begin{document}
\SetRunningHead{T. Enoto et al}{Running Head}
\Received{2007/06/29}
\Accepted{2007/08/17}

\title{Suzaku Observations of Hercules X-1 :\\
Measurements of the Two Cyclotron Harmonics}

\newcommand{\blue}{\textcolor{black}}
\newcommand{\red}{\textcolor{black}}
\newcommand{\referee}{\textcolor{red}}
\newcommand{\green}{\textcolor{green}}


%
 \author{%
   Teruaki \textsc{Enoto},\altaffilmark{1}
   Kazuo \textsc{Makishima},\altaffilmark{1,2}
   Yukikatsu \textsc{Terada},\altaffilmark{2}
   Tatehiro \textsc{Mihara},\altaffilmark{2}\\
   Kazuhiro \textsc{Nakazawa},\altaffilmark{1}
   Tsuyoshi \textsc{Ueda},\altaffilmark{1}
   Tadayasu \textsc{Dotani},\altaffilmark{3}
   Motohide \textsc{Kokubun},\altaffilmark{3}
   Fumiaki \textsc{Nagase},\altaffilmark{3}\\
   Sachindra \textsc{Naik},\altaffilmark{3}
   Motoko \textsc{Suzuki},\altaffilmark{3}
   Motoki \textsc{Nakajima},\altaffilmark{4} 
   and
   Hiromitsu \textsc{Takahashi},\altaffilmark{5}
}
 \altaffiltext{1}{Department of Physics, The University of Tokyo, 7-3-1 Hongo, Bunkyo-ku, Tokkyo 113-0033}
 \email{enoto@amalthea.phys.s.u-tokyo.ac.jp}
 \altaffiltext{2}{Cosmic Radiationn Laboratory, The Institute of Physics and Chemical Research (RIKEN),\\
 2-1 Hirosawa, Wako, Saitama 351-0198}
  \altaffiltext{3}{Institute of Space and Astronautical Science (ISAS), 
Japan Aerospace Exploration Agency (JAXA),\\
Yoshinodai, Sagamihara, Kanagawa 229-8510}
  \altaffiltext{4}{Department of Physics, College of Science and Technology, \\
Nihon University, 8-14,
Kanda-Surugadai 1-chome, Chiyoda-ku, Tokyo 101-8308}
  \altaffiltext{5}{Department of Physical Science, Hiroshima University,\\
 1-3-1 Kagamiyama, Higashi-Hiroshima, Hiroshima 739-8526}


\maketitle

\begin{abstract}
The accretion-powered pulsar  Her X-1 was observed with Suzaku
\red{twice in its main-on state},
on 2005 October 5-6 and 2006 March 29-30, for a net exposure of 30.5 ks and 34.4 ks,
respectively.
In the 2005 and 2006 observations, 
the source was detected at an average 10-30 keV intensity 
of 290 mCrab and 230 mCrab, respectively.
The intrinsic pulse period was measured on both occasions at 1.23776 s by HXD-PIN, 
after barycentric and binary corrections.
The pulse phase-averaged spectra in the energy range above 10 keV 
are well fitted by ``Negative and Positive power-law times EXponential (NPEX)'' model,
multiplied by a fundamental cyclotron resonance scattering feature at $\sim$36 keV
which appears very significantly in the HXD-PIN data.
\red{
The resonance profiles were reproduced successfully 
by the Lorentzian type scattering cross section,
rather than by a Gaussian type alternative.
}
The pulse phase-averaged HXD-GSO data, covering 50-120 keV, are featureless.
However, in a \red{differential} spectrum between the pulse-decay phase and \red{off-pulse} phase,
the second harmonic cyclotron resonance \red{was} detected 
in the GSO data at $\sim$73 keV,
with a depth of 1.6$_{-0.7}^{+0.9}$.
This makes Her X-1 a 6th pulsar with established second harmonic resonance.
Implications of these results are briefly discussed.
\end{abstract}

\section{Introduction} 
Magnetic field strength is 
one of the important fundamental physical parameters of neutron stars.
Their surface magnetic-field strengths can be most accurately determined 
by measuring quantized electron cyclotron resonances,
corresponding to transitions between adjacent Landau levels
which are separated by 
\begin{equation}
\label{crsf}
E_{\rm a} = 11.6 \ B_{12} \cdot (1+z)^{-1} \quad \mathrm{keV},
\label{eq:Ea}
\end{equation}
where $B_{12}$ is the magnetic field strength in units of $10^{12}$ Gauss,
and $z$ is gravitational redshift.
Since this $E_{\rm a}$, with $B_{12}\sim 1$ falls in the X-ray energy range,
accretion-powered X-ray pulsars provide an ideal laboratory where we can 
directly measure $E_a$, and hence $B_{12}$.
Indeed, spectral absorption features at this resonance, 
called cyclotron resonance scattering features (CRSFs),
have so far been detected from more than 15 acceretion-powered X-ray pulsars
\red{
(e.g., \cite{trumper1978,wheaton1978,clark1990,mihara1995,makishima1990,
makishima1999,coburn2002,salvo2004}).
}
Using equation (\ref{crsf}), 
the surface magnetic field strengths of these pulsars 
have been found to cluster in a narrow range of (1-5)$\times 10^{12}$ Gauss
\citep{makishima1999}.

Some of those pulsars with CRSFs exhibit multiple harmonic absorption features.
It was reported that 4U 0115+63 has four harmonics 
\citep{santangelo1999,nakajima2006},
\red{and X0331+63 has up to the third harmonics
\citep{pottschmidt2005,tsygankov2006a,mowlavi2006}.
In addition, there are objects exhibiting 
double (fundamental and second harmonic) CRSFs, 
including Vela X-1 \citep{kreykenbohm1998,makishima1999},
4U1907+09 \citep{cusumano1998,makishima1999}, 
and A0535+26 \citep{kendziorra1994,grove1995}.
}

\red{
Since the fundamental and higher harmonic resonances 
involve somewhat different elementary processes 
(e.g., \cite{alexander+meszaros1991}),
measurements of centroid energies, depths, and widths
of higher harmonics are expected to provide valuable information
on the physics of electron vs. photon interaction in the accretion column.}
Nevertheless, we do not have sufficient understanding as to,
\red{e.g.,} what controls the relative depths 
between the fundamental and second harmonic CRSFs.
Therefore, we still need to enlarge our sample.

The accretion-powered X-ray pulsar Hercules X-1 (Her X-1) is 
one of the most studied objects of this class, over decades
since its discovery in 1972 by Uhuru \citep{giacconi1973}.
At $\sim$35 keV, Her X-1 exhibits a fundamental CRSF,
which is the first CRSF discovered among all pulsars \citep{trumper1978}.
Since then, this $\sim$35 keV CRSF has been studied extensively 
with various X-ray missions, including 
Ginga \citep{mihara1990}, BeppoSAX \citep{dalfiume1998}, RXTE (\cite{gruber2001}, \cite{coburn2002})
and INTEGRAL \citep{klochkov2007}.
Nevertheless, the presence of the second harmonic CRSF in this object has remained controversial.
While the observation with BeppoSAX
obtained evidence of the second harmonic at $E_{\rm a}=72\pm 3$ keV 
in the descending edge of the main pulse peak \citep{salvo2004},
the INTEGRAL data did not confirm it in 2005 observations \citep{klochkov2007}.

 \begin{longtable}{*{5}{c}}
    \caption{Summary of Suzaku observations of Her X-1.}
    \label{table:suzaku_observation_summary}
 \hline \hline
 Epoch        & Start(UT) & End(UT) & Exposure (ks) \footnotemark[$*$] & Position  \\
 \hline
 \endhead
 \hline
 \endfoot
 \hline
 \multicolumn{3}{l}{\hbox to 0pt{\parbox{180mm}{\footnotesize
    \footnotemark[$*$] For the HXD, without dead time correction and elimination of buffer flush intervals.
 }}}
 \endlastfoot
 1 & 2005/10/05 15:12:00& 2005/10/06 10:25:00 & 30.5 & XIS nominal\\ 
 2 & 2006/03/29 18:12:00 & 2006/03/30 15:22:00 & 34.4 & HXD nominal\\
 \end{longtable}

Since the flux of an X-ray pulsar is known to cut off steeply above an energy
of $\sim$1.5$E_a$ \citep{makishima1999},
it is generally not easy to detect the second harmonic CRSF at $\sim 2E_a$.
The Hard X-ray Detector (HXD; \cite{takahashi2007,kokubun2007})
onboard the Suzaku satelite 
has realized high sensitivity over a broad energy band, employing 
Si PIN photo-diodes (hereafter HXD-PIN or briefly PIN) 
and GSO scintillation counts (hereafter HXD-GSO or briefly GSO),
which cover the 10-70 keV and 50-600 keV energy ranges, respectively.
When the X-ray Imaging Spectrometer (XIS; \cite{koyama2007}) is \red{incorporated},
the energy range to be covered expands to three orders of magnitude.
Thanks to the high sensitivity in \red{this} broad energy band,
Suzaku is expected to settle the issue of the second CRSF in Her X-1.
In the present paper, we describe detailed pulse phase-averaged
and phase-resolved spectroscopy of 
Her X-1 made with the Suzaku HXD,
and report on our confirmation of the second harmonic CRSF.

\section{Observations}
Her X-1 has three characteristic periods \citep{nagase1989};
the 1.24 s intrinsic spin period, 
the 1.7 day binary period of the pulsar together with its optical companion HZ Her,
and the 35 day \red{on-off} period 
which is usually attributed to disk precession.
In order to observe a high flux state of Her X-1,
we need to sample so-called main-on phase in the 35 days periodicity,
and avoid the binary eclipses.

We observed Her X-1 twice with Suzaku since its launch.
The first observation was made on 2005 October 5 UT 15:12 through October 6 UT 10:25,
and the second on 2006 March 29 UT 18:12 through March 30 15:22.
These dates were both chosen to observe the main-on phase and to avoid eclipses,
in reference to past observations \citep{zane2004,still2001}.
In both observations, the HXD was operated in the standard mode,
while the XIS employed ``1/8 window'' option to improve the time resolution (to 1 s)
and to avoid event pile up.

\begin{figure}
  \begin{center}
    \FigureFile(80mm,80mm){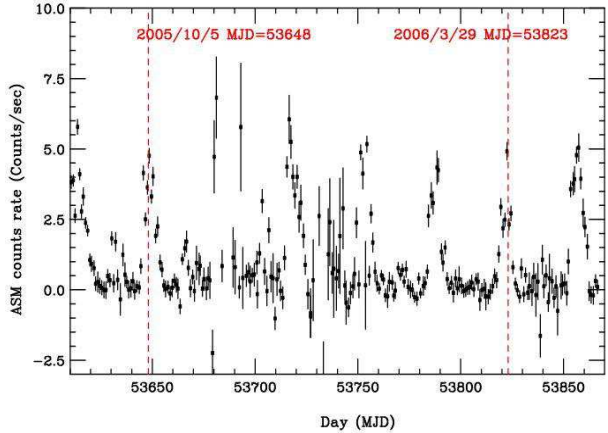}
  \end{center}
  \caption{A 2-10 keV light curve of Her X-1 obtained with the RXTE ASM.
The two Suzaku observations are indicated with red dashed lines.}
\label{./herx1_asm_zoom_pasj2007.eps}
\end{figure}

Figure \ref{./herx1_asm_zoom_pasj2007.eps} shows a long-term light curve 
of Her X-1 obtained by the RXTE ASM.
As indicated there, the two Suzaku observations both sampled the main-on phase as aimed.

\section{Data reduction}
We analyzed the HXD data prepared via version 1.2 pipeline processing.
The data screening criteria we employed are as follows:
(a) the time after passage through the South Atlantic Anomaly 
should be larger than 500 seconds; 
(b) the target object should be above the earth rim by at least 5$\degree$;
(c) geomagnetic cutoff rigidity should be greater than 8 GV $c^{-1}$; and 
(d) the data should be free from ``buffer flash'' \citep{kokubun2007}.
The screenings yielded a net HXD exposure of 30.5 ks and 34.4 ks, 
in the 2005 and 2006 observations, respectively.

\begin{figure*}
  \begin{center}
    \FigureFile(160mm,60mm){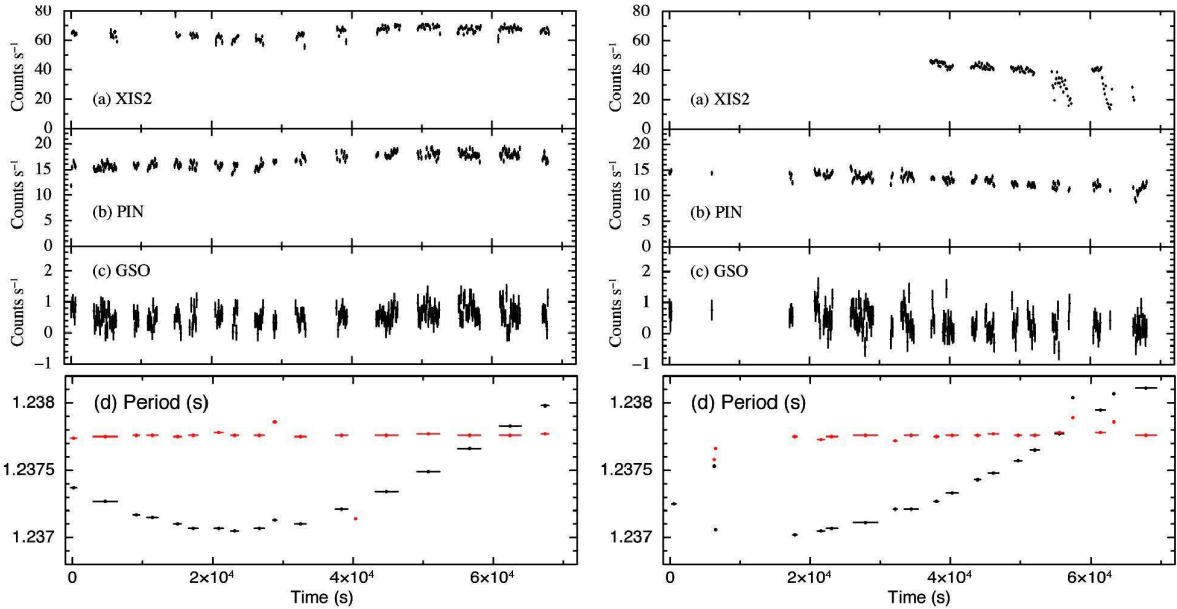}
  \end{center}
  \caption{
    Light curves and period changes of Her X-1 in the 2005 (left) and 2006 (right) observations.
    (a) Background-subtracted 0.4-10 keV XIS2 light curve.
    (b) Background-subtracted and dead-time corrected HXD-PIN light curve in the 10-70 keV energy band.
    (c) The 50-150 keV HXD-GSO light curve, obtained in the same way as the PIN data.
    (d) The pulsation period determined by the folding analysis 
    of the PIN data at each good time interval,
    before (black) and after (red) correcting for the binary motion.
  }\label{fig:xis_pin_gso_lc_pinperiod.eps}
\end{figure*}

Although our main objective is to search the HXD data for the second harmonic CRSF,
we briefly utilize the XIS data as well.
Therefore, we retrieved the XIS data of the two observations,
both processed with version 1.2 pipeline.
Figure \ref{fig:xis_pin_gso_lc_pinperiod.eps}a shows 
the 0.4-10 keV XIS2 light curves.
The XIS background, though included in these light curves, is completely negligible
($\sim$$7\times10^{-2}$ c s$^{-1}$).
The XIS data were not acquired in the former half of the 2006 observation,
due to an operation error.
In the 2006 XIS light curve,
we find a few occasions of intensity decrease.
Since the HXD light curves do not show any corresponding feature,
they are likely to be so-called intensity dips,
observed occasionally from Her X-1 \citep{mihara1991}.

Panels (b) and (c) of figure \ref{fig:xis_pin_gso_lc_pinperiod.eps} are
the background-subtracted and deadtime-corrected light curves
from HXD-PIN (10-70 keV) and HXD-GSO (50-100 keV), respectively.
Thus, both observations were free from binary eclipses.
The source was detected with an average 10-30 keV PIN intensity of 
16.9 cnt s$^{-1}$ and 13.1 cnt s$^{-1}$ in the first and second observations, respectively.

We constructed GSO background (bgd\_d model) 
using a method developed by \citet{fukazawa2007}
for the 2005 and 2006 observations.
We used this GSO background model 
to drive the light curves in figure \ref{fig:xis_pin_gso_lc_pinperiod.eps} 
and to perform standard phase average spectrum analysis in \S 5.1.
Although this GSO background model is available only in relatively coarse energy bins,
finer binnings can be incorporated in the phase-resolved spectroscopy (\S 5.2),
which does not depend on the GSO background model.
To analyze the 2005 data, we utilize
PIN background model called bgd\_a developed by \citet{watanabe2007},
while another model (bgd\_d) developed by \citet{fukazawa2007} for the 2006 PIN data.

\section{Timing Analysis}
Since the XIS data have a time resolution of 1 s,
we conduct the timing analysis only on the HXD data.
The arrival time of each HXD event was corrected for 
the orbital motion effect of the Earth around the Sun,
and that of the satellite around the Earth,
using a Suzaku specific tool {\it aebarycen} \citep{terada2007}
and the object coordinates as 
($\alpha$,$\delta$)=(\timeform{16h57m49.83s}, \timeform{35D20'32''.6}).
The bottom panels (black) of figure \ref{fig:xis_pin_gso_lc_pinperiod.eps} show the pulse period
at each good time interval, 
determined after these corrections by the standard folding analysis
of the 10-70 keV background-inclusive PIN data.
The orbital motion of the pulser is clearly seen.


As a next step, we corrected the event arrival times for the 
orbital delay $\triangle t$ arising in the Her X-1 system,
using a formula as 
\begin{equation}
  \triangle t = \frac{a \sin i}{c} \sin \biggl[ 2 \pi \Big(\frac{t}{P}_{{\rm orb}}  - \phi_0\Big) \biggl].
  \label{binarycorrection}
\end{equation}
Here, $a$ is the semi-mejor axis, $i$ is the inclination, 
$c$ is the speed of light, $t$ is an event time 
(suzakutime with its origin on 2000 January 1th UT 00:00),
$P_{\rm orb}$ is the orbital period, 
and $\phi_0$ is the phase origin.
The values of $(a/c) \ {\rm sin} \ i$ and $P_{\rm orb}$, 
as given in table \ref{tab:system_para_herx1},
were employed, 
while $\phi_0$ corresponding to the suzakutime 0.0 was calculated 
using the phase origin and the values of $P_{\rm orb}$ given by \citet{still2001}.
When we scanned $\phi_0$ over a range of 0.0-1.0,
just for a cross check,
periodograms calculated for sufficiently long time intervals ($\sim$30 ks)
exhibited the strongest contrast correctly 
at $\phi_0\sim$ 0.064.
As presented in figure \ref{fig:xis_pin_gso_lc_pinperiod.eps}(a) in red,
this orbital-delay correction has brought all the instantaneous period measurements 
into a constant value,
$P_0=1.23776$,
in both observations,
with a typical uncertainty of $1\times10^{-5}$ s.
Thus, we quote the intrinsic pulse period of Her X-1 as
$1.23776\pm0.00001$ s,
both on MJD 53646 and MJD 53823.
Any pulse-period difference between the two epochs is 
comparable to this error.

\begin{table}
  \caption{Orbital parameters of Her X-1 employed in the present paper \citep{still2001}.}
  \label{tab:system_para_herx1}
  \begin{center}
    \begin{tabular}{ll}
      \hline \hline
      \multicolumn{1}{c}{Parameter} & Value \\
      \hline
      $ (a/c) \ \mathrm{sin}  \ i $ (s) & 13.19029 \\
      $ P_{\mathrm{orb}}  $ (days)  & 1.7001673 \\  
      $ \phi_0  $\footnotemark[$*$]  & 0.064 \\
      \hline
      \multicolumn{2}{@{}l@{}}{\hbox to 0pt{\parbox{40mm}{\footnotesize
	    \par\noindent
	    \footnotemark[$*$] Orbital phase origin, 
	    defined using equation (\ref{binarycorrection}) at suzakutime 0.0.
	  }\hss}}
    \end{tabular}
  \end{center}
\end{table}


\begin{figure*}
  \begin{center} 
    \FigureFile(160mm,60mm){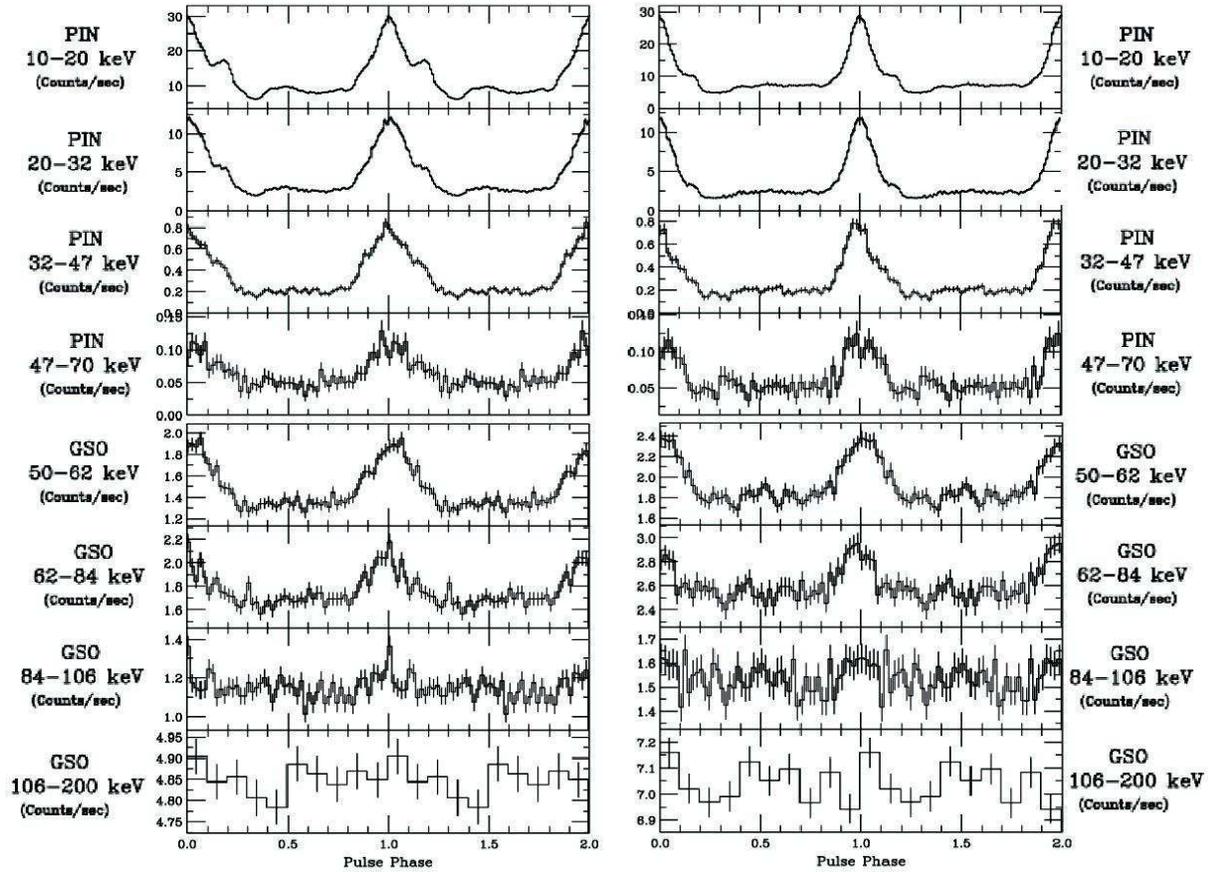}
  \end{center}
  \caption{Energy-sorted and background-inclusive pulse shapes of Her X-1 obtained in the 2005 (left) and 2006 (right) observations. The data were folded by the barycenter- and binary-corrected pulse period,
    $P_0=$1.23776 s. 
    The phase zero ($\phi=0$) is adjusted to the main central peak in the PIN energy band.
  }
  \label{fig:herx1_05_06_plsshp.eps}
\end{figure*}

Figure \ref{fig:herx1_05_06_plsshp.eps} shows 
energy-sorted and background-inclusive pulse profiles of Her X-1
in the two observations.
These were obtained by folding the data at the period of $P_0$.
In both observations,
the pulsation was clearly detected up to 90 keV 
by the HXD.
The measured pulse shapes are typical of the main high state \citep{deeter1998}.
Although they are similar between the two occasions,
a trailing shoulder at phase $\sim$0.2 is more prominent in 2005.

\section{Analysis of the HXD Spectra}

In this section, the HXD spectra obtained in the two observations are analyzed;
the XIS data analysis will be reported elsewhere.
We employ a GSO correction factor, 
which has been introduced by \citet{takahashi_h2007}
to reproduce the Crab spectra by a single power-low model
in the 70-300 keV energy range.
In addition,
we introduce 1\% systematic errors in all spectral fitting analyses,
to reflect  typical uncertainties 
in the current instrumental calibration.
The model normalization is constrained to be the same
between HXD-PIN and HXD-GSO,
while allowed to take different values between the HXS and HXD.
This is because the 1/8 window option of the XIS introduces
some uncertainties in the absolute source flux.

\subsection{Phase-averaged spectra}

\begin{figure}
  \begin{center}
    \FigureFile(80mm,80mm){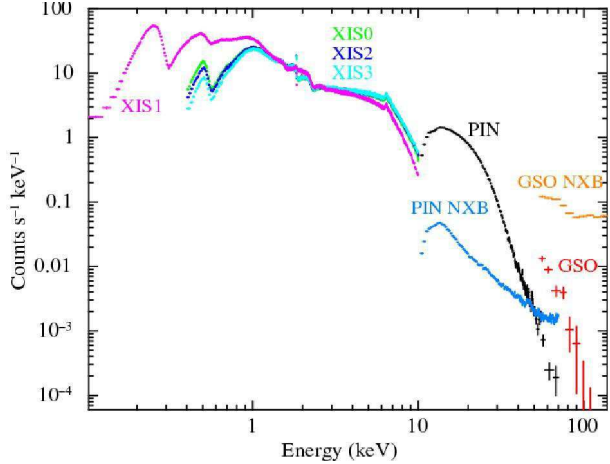}
  \end{center}
  \caption{Background-subtracted and 
    pulse phase-averaged spectra of Her X-1, obtained by the XIS, HXD-PIN, 
    and HXD-GSO.
    Spectra from the 2005 and 2006 observations are summed up.
    The non-Xray background models of PIN and GSO are also shown.
  }
  \label{fig:wideband_sum.eps}
\end{figure}

Figure \ref{fig:wideband_sum.eps} shows 
the background-subtracted 0.1-100 keV spectra of Her X-1,
obtained by the XIS, HXD-PIN and HXD-GSO.
Since the 2005 and 2006 data give very similar spectra,
we co-added them together in figure \ref{fig:wideband_sum.eps}.
The background models (described in \S 3) used for the PIN and GSO data 
are also shown. 
From the XIS data, we subtracted the night earth backgrounds.
After subtracting the background,
the summed data yield an average 1--50 keV flux of 
$6.0 \times 10^{-9}$ erg cm$^{-2}$ s$^{-1}$.
This value was derived by fitting the XIS, HXD-PIN, and HXD-GSO spectra
simultaneously by an empirical model consisting of an NPEX continuum
(to be described later),
and a low-energy blackbody to reproduce the soft X-ray excess.
Assuming an isotropic emission, 
this implies a 1--50 keV luminosity of $3.1\times 10^{37}$ erg s$^{-1}$
at a distance of \red{$D=6.6\pm0.4$} kpc \citep{reynolds1997}.
This luminosity is typical of Her X-1 in the main on state.
\red{
For reference, the first and second observations yielded
the 1--50 keV luminosity of $3.6\times 10^{37}$ erg s$^{-1}$
and $2.5\times 10^{37}$ erg s$^{-1}$, respectively.
}

\begin{figure}
  \begin{center}
    \FigureFile(80mm,56mm){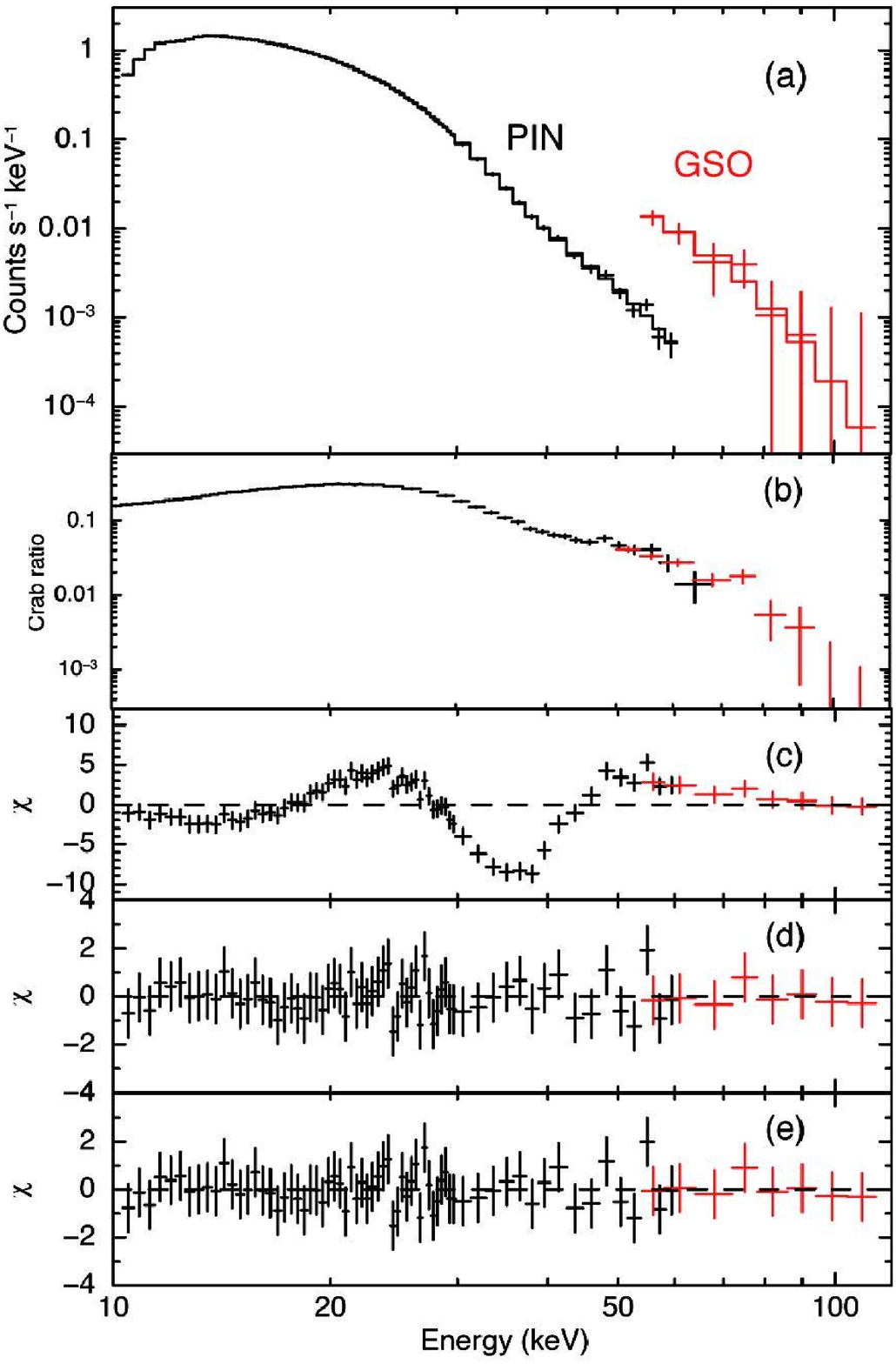}
  \end{center}
  \caption{
    (a) An expanded view of the HXD spectra presented in figure \ref{fig:wideband_sum.eps}.
    Histograms in the figure represent the best fit model consisting of an NPEX continuum and 
    a fundamental cyclotron resonance.
    (b) The same spectra, divided by those of the Crab Nebula.
    (c) Residuals from an NPEX fit.
    (d) Residuals between the data and the best fit model in panel (a).
    (e) Those when a second harmonic CRSF is included.
  }
  \label{fig:phase_ave_spec_fit.eps}
\end{figure}



The HXD spectra in figure \ref{fig:wideband_sum.eps} are expanded 
in figure \ref{fig:phase_ave_spec_fit.eps}a,
and are normalized in figure \ref{fig:phase_ave_spec_fit.eps}b
to the Crab spectra acquired with the HXD 
on 2005 September 15 at the HXD nominal position.
Thus, the source intensity averaged over the two observations was 
$\sim$250 mCrab in the 10-30 keV band.
Independent analyses of the 2005 and 2006 data gave the same quantity 
as $\sim$290 mCrab and $\sim$230 mCrab, respectively.
In the Crab ratio, we observe a broad dip at $\sim$36 keV,
to be identified later with the fundamental CRSF.

To represent the spectral continuum, 
we employed so-called Negative and Positive power-law times EXponential (NPEX) model,
expressed as \citep{mihara1995,makishima1999}
\begin{equation}
  f(E) = (AE^{-\alpha_1} + BE^{+\alpha_2})\times \exp(-E/E_{\rm cut}).
  \label{eq:npex}
\end{equation}
Here, $E$ is the energy, $f(E)$ is the photon number spectrum,
$E_{\rm cut}$ is a cutoff energy,
$A$ and $B$ are normalization factors,
while $\alpha_1>0$ and $\alpha_2>0$ are two photon indices.
We adopt $\alpha_2=2$ for the positive power-law component,
so that the second term represents a Wien hump 
in a saturated inverse Compton spectrum.
First, we fitted the data by this NPEX continuum,
but the fit was not acceptable with a reduced chi-square of $\sim$11.
The residuals, shown in figure \ref{fig:phase_ave_spec_fit.eps}c,
exhibits a strong dip around 36 keV.
This feature, already noted in figure \ref{fig:phase_ave_spec_fit.eps}b,
is interpreted as the CRSF established through the past studies (\S1).

\red{As a next step, 
we multiplied the NPEX continuum by a  factor $e^{-S}$,
where $S$ represents the cyclotron scattering cross section given as
\begin{equation}
  S=\frac{DE^2}{(E-E_a)^2+W^2}\times \left(\frac{W}{E_{\rm a}}\right)^2,
  \label{resonance}
\end{equation}
with $E_{\rm a}$, $D$, and $W$ being the energy, depth, 
and width of the resonance (e.g., \cite{clark1990}, \cite{makishima1999}).
This model has given}
a fully acceptable fit with a reduced chi-square of 0.51.
This value is apparently too small, 
suggesting a slight over-estimation of the systematic error,
but this simply makes our subsequent analysis more conservative.
Histograms in figure \ref{fig:phase_ave_spec_fit.eps}a display this best fit model, 
and figure \ref{fig:phase_ave_spec_fit.eps}d shows residuals 
between the data and the model.
The obtained resonance energy $E_{\rm a} = 35.9_{-0.3}^{+0.3}$ keV,
with its depth $D=1.2_{-0.1}^{+0.1}$ and width $W=12.2_{-1.3}^{+1.5}$,
is generally consistent with the past measurements.
The best-fit parameters are summarized in table \ref{tab:fit_table_phsae_average}.

In the same way, we analyzed the 2005 and 2006 spectra separately,
and obtained the results as presented in table \ref{tab:fit_table_phsae_average}.
\red{Thus, the NPEX times single CRSF model is successful 
on the individual 2005 and 2006 data as well.}
The value of $E_{\rm a}$ is consistent, within errors,
between the two observations,
in which the 10-30 keV counts was different by $\sim$30$\pm10$\%.

In figure \ref{fig:phase_ave_spec_fit.eps}b, 
the Crab ratio suggests a shallow structure at $\sim$70 keV
suggestive of the second harmonic CRSF (\red{\S}1).
In order to quantitatively examine this possibility,
we introduced a model consisting of an NPEX continuum,
multiplied by two CRSF factors both of the form of \red{equation (\ref{resonance})}.
The energy $E_{\rm a2}$ and width $W_2$ of the second CRSF factor 
were fixed at $2E_{\rm a1}$ and $2W_{\rm a1}$, respectively
(with the suffix 1 specifying the parameters of the first CRSF),
while its depth $D_2$ was left free to vary.
However, as presented in figure \ref{fig:phase_ave_spec_fit.eps}e,
this new model did not significantly improve the fit 
to the 2005+2006 spectrum:
the fit chi-square decreased only by 1.2, 
while the degree of freedom changed from 69 to 68.
Therefore, the second harmonic resonance is insignificant,
with the 90\% upper limit being $D_2\le 1.5$.
This is not surprising,
since figure \ref{fig:phase_ave_spec_fit.eps}d reveal little evidence 
for a negative feature at $\sim 2E_{\rm a}=72$ keV.

We repeated the same analysis, namely the NPEX times double CRSF fit,
on the 2005 and 2006 data separately.
The results were essentially the same as before;
the chi-square changed from 45.4 ($\nu=69$) to 44.3 ($\nu=68$) for the 2005 data,
and from 58.4 ($\nu=69$) to 58.1 ($\nu=68$) for the 2006 data,
implying an insignificant improvement in both cases.

\subsection{Phase-resolved spectra}
In many pulsars, the continuum spectra, 
as well as the cyclotron resonance \red{energies and depths}, 
are known to depend significantly on the pulse phase (e.g., \cite{klochkov2007}).
Especially in the phase resolved spectra of Her X-1,
\citet{salvo2004} reported a second CRSF 
at the descending edge of the main pulse peak (\S 1).
We therefore proceed to pulse phase-resolved spectroscopy,
using the 2005 and 2006 data summed up to increase statistics.
In the present case, the phase-resolved analysis has two additional merits:
it allows us to avoid uncertainties in the HXD background models
because we can take direct spectral differences between different pulse phases,
and to use finer bindings of the GSO data than is specified by the current GSO 
background model (\S 2).

\begin{figure*}
  \begin{center}
    \FigureFile(150mm,70mm){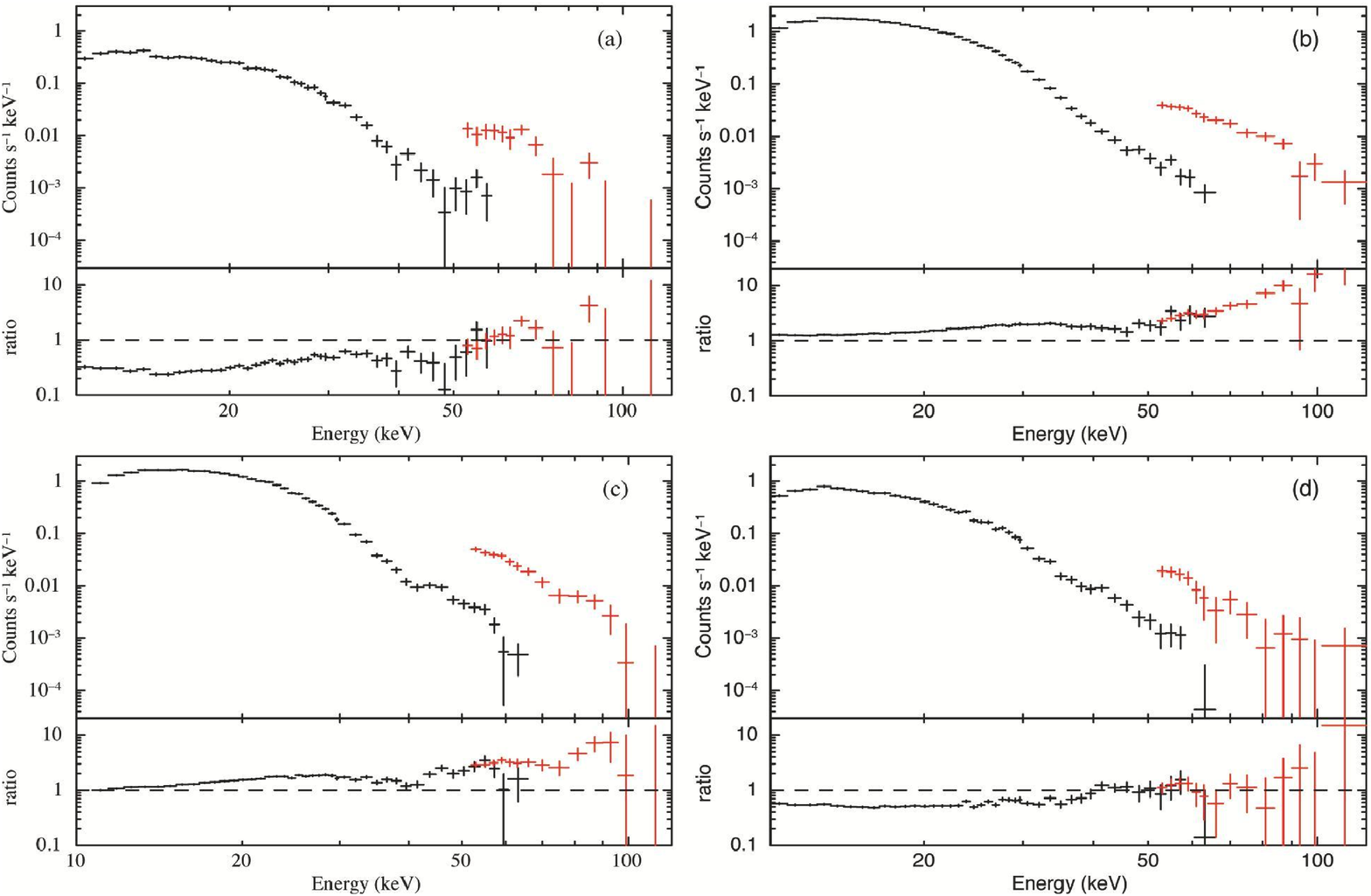}
  \end{center}
  \caption{
    Phase-resolved spectra obtained by subtracting the off-pulse (phase 0.2-0.8) spectrum from
    on-pulse ones;
    (a) $\phi=$0.8--0.9; (b) $\phi=$0.9--1.0; (c) $\phi=$1.0--1.1; (d) $\phi=$1.1--1.2.
    The 2005 and 2006 observations are summed up.
    The bottom half of each panel shows ratios
    of the data, to the best-fit NPEX times single CRSF model
    obtained by the phase-averaged spectra.
  }
  \label{fig:phase-resolved_spectra.eps}
\end{figure*}

Referring to figure \ref{fig:herx1_05_06_plsshp.eps},
we adopted five on-pulse phases as 0.8--0.9, 0.9--1.0, 1.0--1.1, and 1.1--1.2, 
while the off-pulse phase as 0.2--0.8.
The phases 0.8--0.9 and 1.1--1.2 correspond to the soft leading and descending shoulders of the main peak,
while those of 0.9--1.0 and 1.0--1.1 correspond to the leading and 
descending halves of the main hard peak,
respectively.
The phase resolved spectra, obtained by subtracting that of the off-pulse phase, 
are shown in figure \ref{fig:phase-resolved_spectra.eps}.
In these spectra, the PIN and GSO backgrounds are considered to cancel out with
a considerably higher accuracy than is available via the model background subtraction 
($\sim$5\% for PIN and $\sim$2\% for GSO),
because the pulse period of 1.24 s is sufficiently shorter than the time scales of 
typical background variations (minutes to hours),
and because Her X-1 is not bright to cause significant ($>1$\%) 
increases in the HXD dead times.
The bottom half of each panel in figure 6 presents the data, 
divided by the best-fit NPEX times CRSF model determined by the phase-averaged spectra;
the model is meant to provide a rough standard for the phase-resolved spectroscopy.
Thus, some phase-resolved specra are approximately represented 
by the phase-averaged NPEX times CRSF model,
but in general the spectral shapes depends significantly on the phase.

\red{We fitted} the four pulsed-component spectra
by an NPEX continuum and a fundamental CRSF model,
\red{which has given a satisfactory fit to the phase-averaged data.}
Table \ref{fit_table_phsae_resolved} summarizes 
the best fit parameters for individual on-pulse phases,
and figure 7 shows the fit results for the particular phase 
$\phi=1.0-1.1$.
In the case of $\phi$=0.8--0.9, $\phi$=0.9--1.0, and $\phi$=1.1--1.2,
we obtained acceptable fits by this model,
with $\chi^2_{\nu}=$ 1.05, 0.51 and \red{0.62}, respectively.
\red{The fundamental CRSF is highly significant in all these phases,
and the} derived resonance energy, $E_{{\rm a}}=40.6$ keV ($\phi=0.8-0.9$),
$E_{{\rm a}}=39.0$ keV ($\phi=0.9-1.0$),
$E_{{\rm a}}=36.5$ keV ($\phi=1.0-1.1$; see figure 7c),
and 34.8 keV ($\phi=1.1-1.2$), stays within $\sim\pm$10\% of that 
determined with the phase-averaged 2005+2006 spectra.
However, 
as presented in figure 7c,
the model failed to reproduce the $\phi=1.0-1.1$ spectra ($\chi^2_{\nu}=1.67$), 
due to significant negative residuals at $\sim$70 keV in the GSO data which is visible 
even in the ratio between the data and the phase-averaged NPEX times single CRSF model
(figure \ref{fig:phase-resolved_spectra.eps}c).
This can be interpreted as the second harmonic resonance.

We introduced a second CRSF factor, as already performed in \S5.1,
again with the second resonance energy $E_{{\rm a}2}$ fixed at $2E_{\rm a}$.
As shown in table \ref{fit_table_phsae_resolved},
the second CRSF factor hardly improved the fits to the $\phi=0.8-0.9$,
$\phi=0.9-1.0$, and $\phi=1.1-1.2$ spectra.
However, the fit to the $\phi=1.0-1.1$ spectra has been drastically improved
from $\chi^2_{\nu}=1.67$ to $\chi^2_{\nu}=1.07$,
and has become acceptable.
An {\it F}-test indicates that this improvement
being cause by chance is $\sim 1.9 \times 10^{-6}$.
The second harmonic depth is obtainded as $D_2=2.4_{-1.1}^{+0.7}$ (figure 7d).
We therefore conclude that the second harmonic CRSF is
significantly present in the 2005+2006 spectrum of the pulse phase $\phi=1.0-1.1$.


\begin{figure}
  \begin{center}
    \FigureFile(80mm,100mm){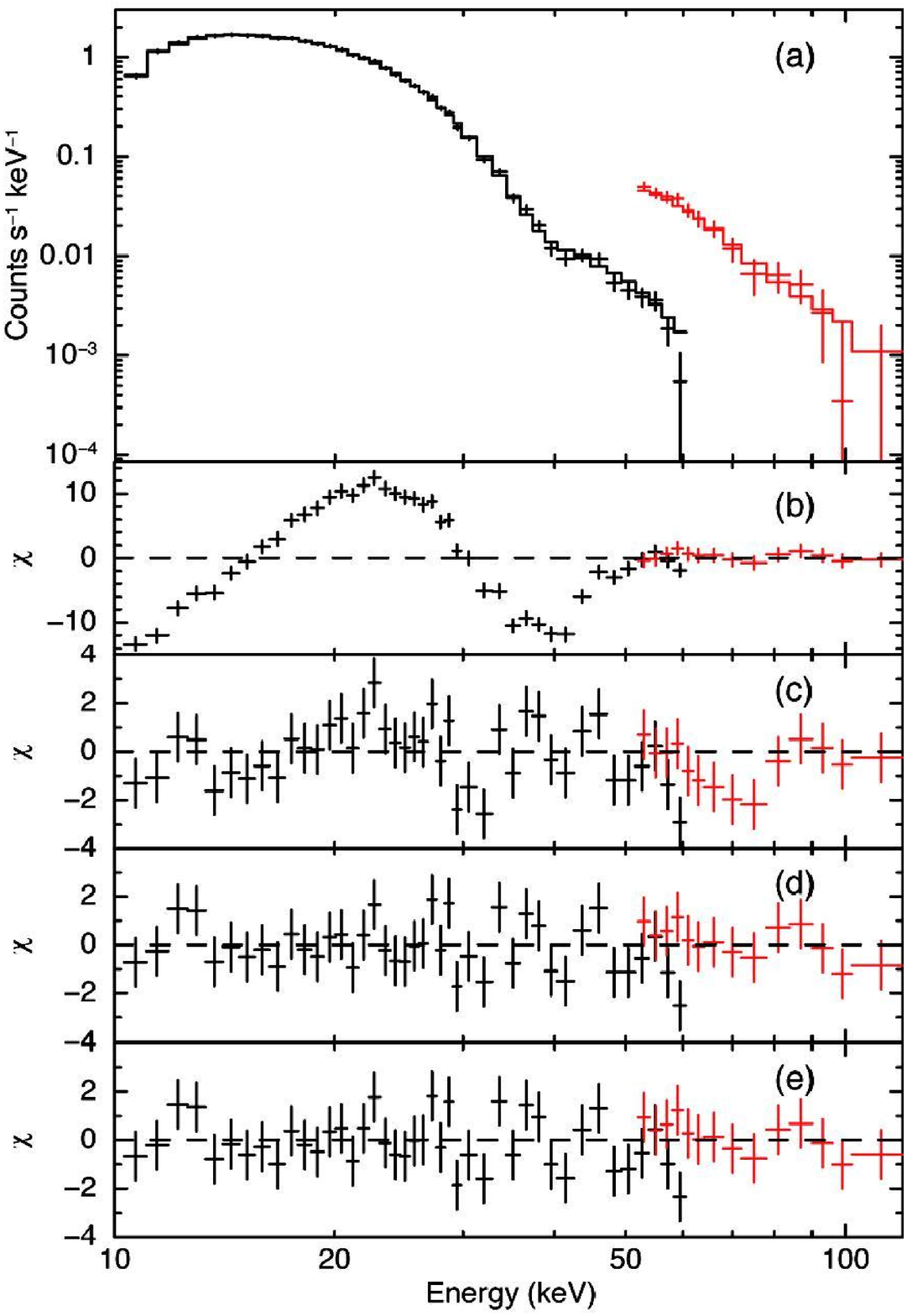}
  \end{center}
  \caption{  (a) Pulsed-component spectra of Her X-1, obtained by subtracting the data of
    the \red{off-pulse} phase ($\phi=0.2-0.8$) from those of the declining phase ($\phi=0.0-0.1$) of 
    the main pulse peak. Data from the 2005 and 2006 observations are summed up.
    The histograms in the figure are the best fit model consisting of an NPEX continuum 
    and double cyclotron resonances; corresponding residuals are displayed in panel (e).
    (b) Residuals when only an NPEX continuum is fitted.
    (c) Fit residuals from the NPEX$\times$CRSF model, with $E_a=36.5$ keV.
    (d) Those with an NPEX and two CRSFs, with $E_{a2}$ fixed at $2E_a$ and $W_2$ fixed at $2W_1$.
    (e) The same as panel (d), but $E_{a2}$ is set free.
  }
  \label{fig:fit_figure2.eps}
\end{figure}

\begin{figure}
  \begin{center}
    \FigureFile(70mm,40mm){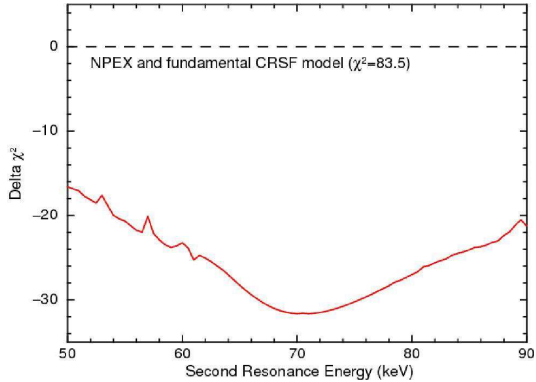}
  \end{center}
  \caption{Decrement in the fit chi-square compared to the NPEX$\times$CRSF fit,
    achieved by adding a second CRSF factor at various energies.
    The same spectra as presented in figure \ref{fig:fit_figure2.eps} are used.
    The width $W_2$ is fixed at $2W_1$. 
    The chi-square of 83.5 obtained by an NPEX and a fundamental CRSF model is 
    also shown.
  }
  \label{fig:sum_0.005_0.105_npex_cyc1_cyc2_steppar_Ea2.eps}
\end{figure}

In order to examine how the present data can constrain the 
second resonance energy $E_{{\rm a}2}$,
we repeated the fitting to the $\phi=1.0-1.1$ spectra 
incorporating the two CRSF factors, but scanning $E_{{\rm a}2}$ over 55-95 keV.
We fixed $W_2$ at $2W_1$.
The behavior of the minimum $\chi^2$, achieved at a given value of $E_{{\rm a}2}$,
is presented in figure \ref{fig:sum_0.005_0.105_npex_cyc1_cyc2_steppar_Ea2.eps}.
Indeed, the chi-square has reached minimum at $\sim$70 keV,
where the second CRSF is expected to appear.
This ensures that the GSO spectrum at this pulse phase has a significant negative 
feature at an energy that is close to $2E_{\rm a}=73.0\pm1.0$ keV
at the same pulse phase.
This NPEX$\times$2CRSFs model yields the second harmonic resonance energy as
$E_{{\rm a}2}=70.2^{+6.9}_{-4.6}$, and its depth $D_2=1.6_{-0.7}^{+0.9}$
(figure \ref{fig:fit_figure2.eps}e).

\blue{
So far,  the second resonance width $W_2$  was fixed to twice that of the fundamental.
When  $W_2$ is made free to vary,
we obtained $W_2=29.3_{-13.9}^{+\infty}$
by constraining $E_{\rm a2}= 2E_{\rm a}$,
or  $W_2 = 21.9_{-10.0}^{+\infty}$ by allowing $E_{\rm a2}$ to vary freely.
Although the upper limit on $W_2$ thus becomes unbound,
the 90\% error range of $W_2$ still includes $2W_1$,
confirming the consistency of our assumption.
}

\section{Discussion}

\subsection{Pulse periods}
We observed Her X-1  twice with Suzaku, in 2005 October and 2006 March,
and obtained high-quality  data with the HXD and XIS
over an extremely broad energy band (0.1--100 keV).
On both occasions, Her X-1 showed X-ray intensity and spectra
typical in the main-on state of its 35-d cycle.

The intrinsic pulse period was determined with HXD-PIN
at $1.23776 \pm 0.00001$ s on both occasions.
\red{
This value agrees with the pulse period history of Her X-1
reported by \citet{staubert2006}.
}

\subsection{The fundamental cyclotron resonance}
In the pulse phase-averaged HXD-PIN spectra
obtained on both occasions,
the fundamental CRSF has been detected clearly  at $\sim 36$ keV.
Using the  NPEX$\times$CRSF spectral model,
we have quantified the resonance parameters
(table~\ref{tab:fit_table_phsae_average},
table~\ref{fit_table_phsae_resolved}).

Based on multiple observations of Her X-1 with RXTE and other missions, 
\citet{gruber2001} and \citet{staubert2007} argued that its fundamental CRSF energy  
exhibited long-term variations, 
possibly related to luminosity changes:
after having stayed at $\sim 35$ keV until 1991,
it  increased after 1991  to nearly 40 keV,
and then it appears to be gradually returning to the previous level. 
The value of $E_{\rm a} = 35.9 \pm 0.3$ keV,
which we obtained  using the pulse phase-averaged 2005+2000 data,
is consistent with the pre-1991 value compiled by \citet{staubert2007},
suggesting that the CRSF energy-change episode, 
whatever the cause be,  has come to an end.

Because  HXD-PIN has a higher sensitivity at $\sim$36 keV than  proportional counters,
and better energy resolution  ($\sim 3$ keV in FWHM) 
than inorganic scintillator instruments,
the present results  provide one of the
best-quality data on the fundamental CRSF of Her X-1.
Importantly, the 10--50 keV PIN spectra bearing the CRSF, 
including  both the phase-averaged
and phase-resolved  ($\phi=1.0-1.1$) ones,
can be reproduced successfully by combining
the NPEX continuum of equation (\ref{eq:npex})
with the scattering cross section of equation (\ref{resonance});
the CRSF profile has been resolved with a reasonable accuracy.
We therefore reconfirm the appropriateness of this modeling,
which has been used in many of the past studies of CRSFs
including those with Ginga \citep{clark1990, mihara1990, mihara1995, makishima1999}, 
RXTE \citep{nakajima2006}, INTEGRAL \citep{tsygankov2006a},
and Suzaku \citep{terada2006}.

As an  exercise,
we replaced the cyclotron absorption factor 
$\exp(-S)$ ($S$ referring to equation~\ref{resonance})
with a Gaussian absorption factor 
used in some other studies (e.g, \cite{staubert2007,klochkov2007});
$\exp\left[-a \exp\{(E-E_{\rm c})^2/2\sigma^2\}\right]$,
where  $a$, $E_{\rm c}$,  and $\sigma$ are free parameters.
When this factor is applied to the NPEX continuum,
the fit to the phase-averaged  HXD (2005+2006) spectra
worsened  to $\chi^2_\nu = 0.70$ ($\nu=69$),
from that obtained using  equation~(\ref{resonance})
($\chi^2_\nu = 0.52$ for $\nu=69$; table~\ref{tab:fit_table_phsae_average}).
Although this fit is still acceptable
(due possibly to an over-estimated systematic error),
the Gaussian absorption  has failed to reproduce 
the $\phi=1.0-1.1$ spectra  with  \red{$\chi^2_\nu = 2.$72 ($\nu=48$),}
even when two of them (for the two harmonic CRSFs) were incorporated.
Therefore, at least for  Her X-1,
the Lorentzian-like form of equation~(\ref{resonance}),
which is based on the classical cross section 
of cyclotron resonance \citep{clark1990},
is considered more appropriate than the alternative Gaussian cross section.

The above argument may need an important remark:
the very deep fundamental CRSF of the transient pulsar X0331+53 (V0332+53), 
at about 28 keV, cannot be described adequately 
if using a single form of equation (\ref{resonance})
\citep{makishima1990, nakajima2006, tsygankov2006a}.
Instead, a nested pair of such  absorption factors with different widths 
(e.g., in terms of Gaussians; \cite{pottschmidt2005}) may be needed.

\subsection{Possible origins of the resonance width}
\label{subsec:widthorigin}

What is the origin of the relatively large width of CRSFs,
which is generally expressed as
\begin{equation}
W = (0.2-0.5) E_{\rm a}
\label{eq:W1Ea}
\end{equation}
 \citep{makishima1999,nakajima2006}?
It has often been argued
that $W$ is determined mainly by thermal Doppler effects in the accretion column,
including those associated with  electron motion
along the magnetic field lines  (e.g., \cite{dalfiume1998,cusumano1998,coburn2002}).
However, the failure  of  the Gaussian optical depth
to reproduce the present Suzaku data casts doubt on this interpretation.
Furthermore, as  pointed out  by  \citet{makishima1999},
the values of $W$ of various X-ray pulsars 
often exceed the expected thermal broadening.
According to \citet{nakajima2006},
$W_1$ of the transient pulsar X0115+63 increased 
by a factor of $\sim 5$ as its luminosity changed from 
$2 \times 10^{36}$ erg s$^{-1}$ to $5 \times 10^{37}$ erg s$^{-1}$,
but the temperature of the emission region,
$\propto E_{\rm cut}$ of equation (\ref{eq:npex}),
increased  by only $\sim 20\%$.
A similar luminosity dependence of $W$ may be
visible even between the present two observations.
These observed changes in $W$ 
also argue against the thermal broadening scenario.

In explaining  $W$, an obvious alternative is
pulse phase-dependent changes in $E_{\rm a}$.
However, in our table~\ref{tab:fit_table_phsae_average}
and table~\ref{fit_table_phsae_resolved},
$W_1$ is not necessarily smaller in phase-resolved spectra 
(except in  $\phi=1.1-1.2$)  than in the phase-averaged ones.
Furthermore, the superposition of narrow features with different
$E_{\rm a}$ over the pulse phase would not produce
wide Lorentzian-like wings.
Therefore, this interpretation is not likely, either.

In classical electrodynamics,
the width $W$ in equation (\ref{resonance}) represents
damping effects on the gyrating electrons.
In quantum mechanics,
the resonance width in this formula generally reflects 
the finiteness of the life time of excited states, 
as   $W \sim h\Lambda$, through uncertainty principle:
here, $\Lambda$ is transition rate of electrons out of  the relevant excited state, 
and $h$  is the Planck constant.
This $\Lambda$, in turn, 
is determined by three competing de-excitation processes;
namely, collisional, spontaneous radiative, and stimulated (induced).
Among them, the spontaneous transition rate $\Lambda_{\rm rad}$
(so-called Einstein's A-coefficient) is given as
\[
 \Lambda_{\rm rad} \sim 3.8 \times 10^{15} B_{\rm 12} ~{\rm s}^{-1}
\]
\citep{meszaros1992}, 
so that the resonance width due to spontaneous emission,
namely natural width, becomes
\begin{equation}
W_{\rm nat}  \sim  h \Lambda_{\rm rad} = 15 B_{12} ~ {\rm eV} = 1.3 \times 10^{-3}  E_{\rm a}
\label{eq:naturalwidth}
\end{equation}
where the final form employs equation (\ref{eq:Ea})
neglecting the gravitational redshift $z$.
This is still inadequate to explain equation (\ref{eq:W1Ea}).
A brief calculation shows  that the collisional de-excitation is even less effective.

Let us finally consider the stimulated emission,
of which the rate is given as $\Lambda_{\rm st}=\lambda_{\rm st} J_{\rm a}$,
where $\lambda_{\rm st}$ is so-called Einstein's B coefficient, 
and $J_{\rm a}$ is the radiation energy flux per unit photon frequency 
at the resonance energy $E_{\rm a}$;
this effect was considered by \citet{alexander+meszaros1991}.
As estimated very crudely in Appendix,
$\Lambda_{\rm st} $ in Her X-1 could be some two orders of 
magnitude larger than $ \Lambda_{\rm rad}$.
Then, from equation  (\ref{eq:naturalwidth}),
the resonance width due to stimulated emission, 
$W_{\rm st} = W_{\rm  nat} (\Lambda_{\rm st} /  \Lambda_{\rm rad}) $,
could be a considerable fraction of $E_{\rm a}$,
and might potentially provide an explanation to the observed width.
An advantage of this picture is
that we can naturally explain the observed positive dependence 
of $W$ on the X-ray luminosity \citep{nakajima2006},
because we expect $W_{\rm st} \propto \Lambda_{\rm st} \propto J_{\rm a}$.
On the other hand, a caveats is 
that equation (\ref{eq:detailed_balance}) in Appendix predicts the $W/E_{\rm a}$ ratio 
to decrease steeply toward the higher-field objects,
although such a scaling  is not necessarily supported by observation. 
Furthermore, it is not obvious if the absorption feature  can be
really produced in such a condition as $\Lambda_{\rm st} /  \Lambda_{\rm rad}\gg 1$,
which would usually emhance emission.
We leave this intriguing issue to further studies.

\subsection{The second harmonic resonance}
The pulse phase-averaged HXD spectra 
did not require the second harmonic resonance,
beyond a 90\%  upper limit of  $D_2\le 1.5$.
However, in the pulse phase-resolved spectra
which cover the descending edge of the main pulse ($\phi=1.0-1.1$),
we have successfully detected  the second CRSF with $D_2=2.4_{-1.1}^{+0.7}$
(under the constraints of $E_{\rm a2}=2E_{\rm a}$ and $W_2 = 2W_2$).
Its energy, when allowed to vary, 
becomes  $E_{\rm a2}=70.2 ^{+6.9}_{-4.6}$,
yielding $E_{\rm a2}/E_{\rm a}=1.9\pm 0.2$
which  satisfies the 1:2 harmonic ratio.
Here,  the value of $E_{\rm a}$ can be taken either as the phase-averaged value,
or that  at this particular pulse phase, because they agree within 2\%.

As evidenced by the large chi-square decrements 
(figure~\ref{fig:sum_0.005_0.105_npex_cyc1_cyc2_steppar_Ea2.eps}),
the present detection of the second harmonic feature is statistically highly significant.
The feature cannot be artifacts due to 
systematic errors in  the  background  subtraction (\S5.2),
since background systematics can be neglected
in these  ``on-pulse minus off-pulse" spectra.
Another concern is the uncertainty in the GSO response,
toward lower energies \citep{kokubun2007}.
We confirmed that  the second harmonic resonance remains significant
even when we discard the GSO energy ranges below 60 keV,
although errors on the second resonance parameters increase.
We therefore conclude that the detected feature is real.

\begin{figure}
  \begin{center}
    \FigureFile(70mm,50mm){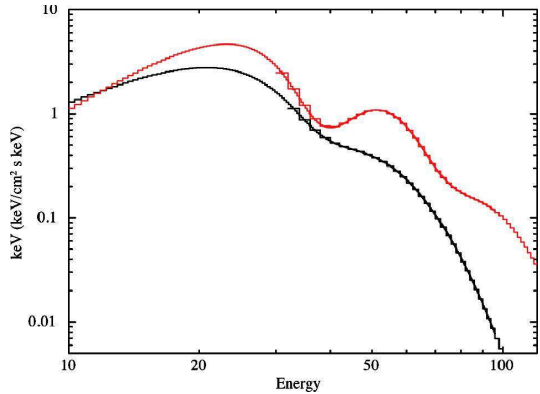}
  \end{center}
  \caption{
The best-fit $\nu f \nu$ model spectra.
The black line shows the best-fit NPEX continuum multiplied by a fundamental CRSF
determined using the 2005+2006 pulse phase-averaged spectrum.
The red line shows the best-fit NPEX continuum multiplied by double CRSFs
specified by a 2005+2006 pulse phase-resolved spectrum ($\phi=1.0-1.1$).
     }
 \label{fig:nufnu_eemodel.eps}
\end{figure}

Evidently, HXD-GSO played a major roll in detecting the second CRSF,
which lies outside the HXD-PIN energy range.
Nevertheless, the PIN data also contributed to its  detection,
because the inclusion of the second resonance factor has
considerably decreased fit residuals in the PIN energy range
as well (figure~\ref{fig:fit_figure2.eps}c, d, e).
In figure \ref{fig:nufnu_eemodel.eps},
we show the inferred best-fit model spectrum
for phase-averaged with a fundamental CRSF
and for phase-resolved with double CRSFs.

The second CRSF of Her X-1 was first reported by  BeppoSAX,
from a main-on state observation in  October 2000  \citep{salvo2004}.
It appeared in the descending edge of the main pulse peak,
with the  resonance energy at $E_{\rm a2} \sim 72 \pm 3$ keV 
and its width  $W_2 \sim 11 \pm 1$ keV.
Our results agree  with those from BeppoSAX,
with respect to the  resonance energy,
as well as the particular pulse phase where it becomes significant.
These results hence make Her X-1 
a 6th pulsar with  the second harmonic resonance.
The negative detection of this feature with  INTEGRAL,
in another main-on state observation conducted 
in July-August 2005 \citep{klochkov2007},
is probably due to insufficient statistics.

\begin{figure}
  \begin{center}
    \FigureFile(70mm,50mm){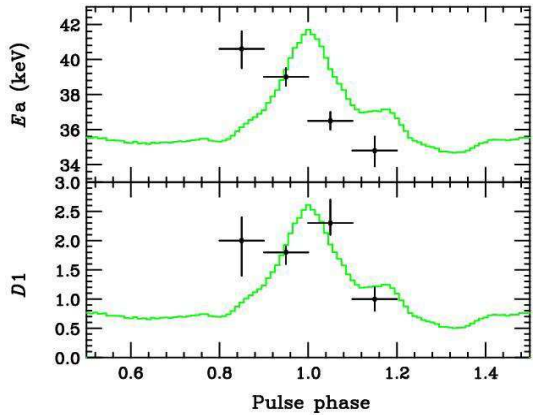}
  \end{center}
  \caption{
    (top) Fundamental CRSF energies presented as a function of the pulse phase,
    obtained by subtracting the interpulse spectrum 
    from on-pulse ones using the 2005+2006 data.
    The 10--20 keV pulse profile (also summed over 2005 and 2006) is superposed.
    (bottom) Fundamental CRSF depths presented in the same manner.
     }
 \label{fig:phase_parameters_energy_depth.eps}
\end{figure}

The present results suggest
that the second harmonic resonance is a rather 
common feature of accreting X-ray pulsars,
particularly when they are luminous \citep{nakajima2006}.
Even though the cross section of photon vs. electron interaction
in a strong magnetic field is much larger at $E_{\rm a}$ than at $2E_{\rm a}$,
the  two-photon effect by \citet{alexander+meszaros1991} ensures 
that the second harmonic feature can appear as strong as the fundamental.
That is, the fundamental resonance acts essentially as a photon scattering process, 
due to  very short lifetime of the excited state.
In contrast, the second resonance is expected to act as a pure absorption,
because an electron excited by two Landau levels will return 
to the ground state by emitting two photons with energy $\sim E_{\rm a}$;
these photons fill the fundamental feature, and make it shallower.

\subsection{Pulse-phase dependence of the resonances}

Figure \ref{fig:phase_parameters_energy_depth.eps} shows 
the measured $E_{\rm a1}$ and $D_1$ as a function of the pulse phase;
numerical values are given in table \ref{fit_table_phsae_resolved}.
The results generally agree with the 
INTEGRAL measurements \citep{klochkov2007}.
Thus, in the particular pulse phase ($\phi=1.0-1.1$) 
where the second CRSF was detected,
the fundamental resonance is also very deep
showing a large value of  $D_1$,
together with a relatively low $E_{\rm a1}$ (though not the lowest).
Then, how this particular pulse phase relate to the rotational phase of the pulsar?

Although the rotational modulation of $E_{\rm a1}$ 
does not allow straightforward explanation,
we can think of a simple physical effect related to it;
the luminosity-dependent change in $E_{\rm a1}$.
In the transient pulsars X0115+63 \citep{mihara2004,nakajima2006}
and X0331+53 \citep{tsygankov2006b,nakajima_dron2006},
$E_{\rm a1}$ was found to decrease as the luminosity becomes high
enough (several times $10^{37}$ erg s$^{-1}$ or higher).
This is presumably because the accretion column then gets taller,
so the resonance photosphere, 
as seen from  end-on directions,
appears at a higher altitude in the column 
where the dipole field intensity is lower  \citep{mihara2004}.
If this effect is also taking place in Her X-1,
we  expect to observe a relatively low value of $E_{\rm a}$
when looking onto one of the accretion poles.
In contrast, when viewing an accretion column relatively side on,
we will sample various heights along it,
and to measure a higher value of $E_{\rm a}$.

Another candidate mechanism to produce the phase-dependent change in $E_a$ is 
the angle-dependent relativistic effect pointed out by \citet{harding1991}.
For $E_a\sim$36 keV, this effect makes the exact value of $E_a$ to decrease by $\sim$3\%
at the pole-on phase than at the side-on phase.
Although this shift is small, 
it works in the same sense as the previous one. 

In addition to the above two mechanism,
yet another effect may also produce the phase dependence of $E_{\rm a}$; 
namely, bulk Doppler effect in the accretion column.
Since the accretion flow onto a pulsar 
has a typical velocity of $v_{\rm fl} \sim c/3$,
we expect the post-shock plasma in the radiating accretion column
to have a bulk flow of $\sim v_{\rm fl}/4 \sim 0.1 c$,
at least just beneath the standing shock surface.
Then, the longitudinal Doppler shift associated with 
this bulk flow will make $E_{\rm a}$ lower by $\sim 10$\% 
when we look onto the column,
while the effect will vanish at side-on phases.
Actually, \citet{terada2004} invoked such Doppler shifts in 
accretion columns of magnetized white dwarfs,
and successfully explained the unusually strong atomic emission lines
observed from several objects of that kind.
In X-ray pulsars, 
this mechanism is expected to work
in the same sense as the above two effects,
thus enhancing one another.

From these considerations, we tentatively conclude 
that the descending half of the main pulse,
where $E_{\rm a}$ is relatively low,  corresponds to the phase 
where we are looking end-on into one of the two accretion poles.
This agree with theoretical pulse-profile decomposition by \citet{leahy2004}.
According to this work,
the descending half corresponds to the pole-on phase of one pole;
the pulse profile becomes asymmetric with respect to this  phase,
due to an addition of fan-beam emission from the other pole
which reaches us after affected by gravitational light bending.
If this phase assignment is correct, we are to conclude
that the second harmonic feature becomes most prominent
when we are observing one pole from and end-on aspect.
Further comparison with other pulsars,
including in particular X0115+63 and X0331+53,
would be of great value,
though beyond the scope of the present paper.

\section{Conclusion}
In this paper, we analyzed the HXD data from two Suzaku observations of Her X-1.
The fundamental resonance was clearly detected at $\sim$36 keV 
in the pulse phase-averaged and phase-resolved spectra.
The Lorentzian-like form of the resonance was found to be appropriate 
than the alternative Gaussian cross section.
The second resonance feature, 
though absent ($D_2\le 1.5$) in the phase-averaged spectra,
was detected at $\sim$73 keV in the pulse phase-resolved spectra
at the descending edge of the main peak.

The authors are grateful to all the members of the
Suzaku Science Working group,
for their help in spacecraft operation,
instrumental calibration,
and data processing.

\red{
\section*{Appendix: Estimation of induced emission}
Assuming a local thermal equilibrium,
detailed balance relates  the Einstein's A and B coefficients as
$\lambda_{\rm st} = \Lambda_{\rm rad} (hc)^2/(2 E_{\rm a}^3)$,
where the quantities are defined in subsection \ref{subsec:widthorigin}.
Then, the stimulated transition rate,
relative to the spontaneous one, can be written as
\begin{equation}
\Lambda_{\rm st} /  \Lambda_{\rm rad} = (hc)^2 J_{\rm a}/2E_{\rm a}^3~~.
\label{eq:detailed_balance}
\end{equation}
}

\red{
Let us next estimate the spectral flux density $J_{\rm a}$ at 36 keV.
The observed flux density per unit energy at $E_{\rm a}$,
{\em before} the CRSF factor is applied,
is derived from our best fit model to the phase-resolved spectra
as $3.6 \times 10^{-10}$ erg s$^{-1}$ cm$^{-2}$ keV$^{-1}$.
Multiplying this by $4\pi D^2$, 
where $D=6.6$ kpc (subsection 5.1) is the distance,
we obtain a monochromatic luminosity at $E=E_{\rm a}$
as $l=1.8 \times 10^{36}$ erg$^{-1}$ s$^{-1}$ keV$^{-1}$.
Further assuming that the two polar emission regions of Her X-1 
can be approximated each by a hemisphere of radius $r \sim 100$ m, 
we obtain
\begin{eqnarray*}
J_{\rm a} &=&  \frac{l}{4\pi r^2} \\
&=&  3.9 \times 10^{20} 
  \left( \frac{r}{\textrm{100m} } \right)^{-2} 
  \left( \frac{E_{\rm a}}{36~\textrm{keV}} \right)^{-3}
 \textrm{keV} ~\textrm{s}^{-1} ~\textrm{cm}^{-2} ~\textrm{Hz}^{-1} 
\end{eqnarray*}
after transforming units from erg  keV$^{-1}$ to keV Hz$^{-1}$.
This finally yields
\[
\Lambda_{\rm st} /  \Lambda_{\rm rad} \sim 62 
  \left( \frac{r}{\textrm{100m} } \right)^{-2} 
  \left( \frac{E_{\rm a}}{36~\textrm{keV}} \right)^{-3}~~.
\]
}


 \begin{longtable}{*{11}{c}}
    \caption{Best-fit parameters of the phase averaged spectra of Her X-1.}
     \label{tab:fit_table_phsae_average}
 \hline \hline
 model$^*$     & $A^{\dagger}$ & $B^{\dagger}$ & $\alpha_1$ & $E_{\rm cut}$ & $E_{\rm a}$ & $D_1$ &  $W_1$    & $E_{{\rm a}2}$ & $D_2$ &  $\chi^2_{\nu}$($\nu$)\\
 & ($\times 10^{-2}$) & ($\times 10^{-2}$)  &  & (keV) &   (keV)  & & (keV)  & (keV) &  &  \\
 \hline
 \endhead
 \hline
 \endfoot
 \hline
 \multicolumn{3}{l}{\hbox to 0pt{\parbox{150mm}{\footnotesize
    \footnotemark[$*$] NP represents NPEX continum model, C1 represents a foundamental CRSF.
 C12 represents foundamental and second CRSFs with a second resonance energy fixed at the
twice as a foundamental energy, and with the width of a second resonance fixed at the
twice as a foundamental one.
 C1$\times$C2 represents foundamental and second CRSFs with both resonance enrgy free,
but with the width of a second resonance fixed at the twice as a foundamental one.\\
\footnotemark[$\dagger$] Referring to equation (\ref{eq:npex}), and defined at 10 keV in
unit of counts keV$^{-1}$ cm$^{-2}$ s$^{-1}$.
 }}}
 \endlastfoot
2005 &   &  &  &  &  &  &  &  &  &   \\
\hline
{\small NP } & $<0.1$ & 0.1 & -- & 5.6 & -- & -- & -- & -- & -- & 8.52 (72)   \\
{\small NP$\times$C1} & 2.8 & 3.4 & 0.4$_{-0.4}^{+0.7}$ & 7.3$_{-0.6}^{+0.6}$  &  36.0$_{-0.4}^{+0.6}$ & 1.5$_{-0.2}^{+0.2}$ & 15.3$_{-2.1}^{+2.0}$ & -- & -- & 0.66 (69)   \\
{\small NP$\times$C12} & 3.0 & 2.0 & 0.1$_{-0.1}^{+0.6}$ & 8.7$_{-2.0}^{+2.8}$ &  35.8$_{-0.6}^{+0.6}$  & 1.8$_{-0.2}^{+0.2}$ & 16.4$_{-1.6}^{+1.1}$ & -- & 1.0$_{-0.5}^{+1.0}$ & 0.65 (68)  \\
\hline

2006 &  &  &  &  &  &  &  &  &  &  \\
\hline
{\small NP} & $<0.1$ & 7.6 & --- & 5.0 & -- & -- & -- & -- & -- & 6.9 (72)   \\
{\small NP$\times$C1} & 0.9 & 4.6 & 1.2$_{-1.2}^{+2.0}$ & 5.8$_{-0.3}^{+0.5}$  & 35.5$_{-0.5}^{+0.5}$ & 0.9$_{-0.1}^{+0.2}$ & 9.6$_{-1.3}^{+1.5}$ & -- & -- & 0.85 (69)  \\
{\small NP$\times$C12} & 0.8 & 4.6 & 1.1$_{-1.1}^{+1.9}$ & 5.8$_{-0.3}^{+2.3}$  & 35.3$_{-0.6}^{+0.3}$ & 0.9$_{-0.1}^{+0.2}$ & 9.3$_{-1.3}^{+1.6}$ & -- & 0.2$_{-0.2}^{+0.9}$ & 0.85 (68)  \\
\hline
sum &  &  &  &  &  &  &  &  &  &   \\
\hline
{\small NP} & $<0.1$ & 9.7 & -- & 4.9        & -- & -- & -- & -- & -- & 10.7 (72)  \\
{\small NP$\times$C1} & 2.0 & 4.2 & $0.5_{-0.5}^{+0.9}$  & $6.4_{-0.4}^{+0.5}$ &  $35.9_{-0.3}^{+0.3}$  & $1.2_{-0.1}^{+0.1}$  & $12.2_{-1.3}^{+1.5}$           & -- & -- & 0.52 (69)  \\
{\small NP$\times$C12} & 2.5 & 2.5 & $0.0_{-0.0}^{+0.5}$ & $7.4_{-1.2}^{+6.3}$ & $35.6_{-0.5}^{+0.4}$  & $1.4_{-0.1}^{+0.1}$  & $13.6_{-1.1}^{+1.4}$           & -- & $1.0_{-0.4}^{+0.5}$ & 0.51 (68)  \\
 \end{longtable}

 \begin{longtable}{*{11}{c}}
    \caption{Best-fit parameters of the phase-resolved 2005+2006 spectra of Her X-1.}
     \label{fit_table_phsae_resolved}
 \hline \hline
  model$^*$      & $A^{\dagger}$ & $B^{\dagger}$ & $\alpha_1$ & $E_{\rm cut}$ & $E_{\rm a}$ & $D_1$ &  $W_1$    & $E_{{\rm a}2}$ & $D_2$ &  $\chi^2_{\nu}$($\nu$)\\
   & ($\times 10^{-2}$) & ($\times 10^{-2}$)&  & (keV) &   (keV)  & & (keV)  & (keV) &  &  \\
 \hline
 \endhead
 \hline
 \endfoot
 \hline
 \multicolumn{3}{l}{\hbox to 0pt{\parbox{150mm}{\footnotesize
    \footnotemark[$*$] Abbreviations and the definition of parameters are the same as in table \ref{tab:fit_table_phsae_average}.
 }}}
 \endlastfoot
\hline
$\phi=$0.8-0.9  &  &  &  &  &  &  &  &  &  & \\
 {\small NP$\times$C1}  & 0.9 & 6.6 & 3.2$_{-1.2}^{+1.5}$ & 8.0$_{-0.5}^{+0.8}$ &
  40.6$_{-1.0}^{+1.1}$ & 2.0$_{-0.4}^{+0.6}$ & 7.5$_{-1.8}^{+3.4}$ & -- & -- & 1.05 (50) \\
 {\small NP$\times$C12} & 0.8 & 2.6 & 2.2$_{-0.9}^{+1.2}$ & 12.6$_{-4.1}^{+8.1}$     &
40.4$_{-1.3}^{+1.3}$ & 2.6$_{-0.7}^{+0.5}$ & 10.9$_{-3.3}^{+2.7}$  & -- & 3.1$_{-2.6}^{+2.4}$ & 0.99 (49) \\

\hline
$\phi=$0.9-1.0  &  &  &  &  &  &  &  &  &  & \\
 {\small NP$\times$C1} & 2.1 & 3.7 & 2.2$_{-1.2}^{+1.4}$ & 7.8$_{-0.3}^{+0.4}$  &
  39.0$_{-0.5}^{+0.5}$ & 1.8$_{-0.1}^{+0.2}$ & 11.6$_{-1.3}^{+2.0}$ & -- & -- & 0.51 (50) \\
 {\small NP$\times$C12} & 2.4 & 2.1 & 1.1$_{-0.9}^{+0.9}$ & 10.1$_{-2.3}^{+4.3}$  &
  38.6$_{-0.8}^{+0.7}$ & 2.2$_{-0.1}^{+0.3}$ & 13.8$_{-2.0}^{+1.5}$ & -- & $1.4_{-1.4}^{+1.5}$ & 0.49 (49) \\

\hline
$\phi=$1.0-1.1  &  &  &  &  &  &  &  &  &  & \\
{\small NP$\times$C1} & $<0.1$ & 4.4 & $\cdots$ & 7.3     &
  37.4   & 1.9         &  8.4     & --     & --         & 1.67 (50)  \\
{\small NP$\times$C12}  & 0.7 & 2.3 & 0.5$_{-0.5}^{+13.1}$ & 10.3$_{-1.9}^{+1.3}$     &
  36.5$_{-0.5}^{+0.5}$ & 2.3$_{-0.4}^{+0.2}$ & 11.2$_{0.9}^{+1.0}$  & -- & 2.4$_{-1.1}^{+0.7}$ & 1.07 (49)\\
 {\small NP$\times$C1$\times$C2} & 0.4 & 3.0 & 1.7$_{-1.7}^{+14.4}$ & 8.9$_{-0.9}^{+3.3}$     &
  36.5$_{-0.2}^{+0.4}$  & 2.2$_{-0.2}^{+0.4}$ & 10.2$_{-0.9}^{+2.0}$ & 70.2$_{-4.6}^{+6.9}$ & 1.6$_{-0.7}^{+0.9}$  & 1.08 (48) \\

\hline
$\phi=$1.1-1.2  &  &  &  &  &  &  &  &  &  & \\
{\small NP$\times$C1} & 1.9 & 1.7 & 1.2$_{-1.1}^{+1.4}$ & 6.6$_{-0.3}^{+0.4}$    &
  34.8$_{-0.8}^{+0.9}$    & 1.0$_{-0.2}^{+0.2}$         &  4.5$_{1.7}^{+2.6}$    &     -- & --        & 
0.62 (50) \\
 {\small NP$\times$C12} & 1.9 & 1.7 & $1.2_{-1.1}^{+1.4}$  & $6.6_{-0.3}^{+2.4}$    &
 34.8$_{-0.8}^{+0.8}$    & $1.0_{-0.2}^{+0.2}$ & 4.5$_{2.1}^{+2.6}$ &  -- & $0.0_{-0.0}^{+1.1}$        & 0.63 (49) \\

 \end{longtable}

\end{document}